\documentclass[12pt,fleqn]{book} 
\usepackage{tcolorbox}
\usepackage[utf8]{inputenc}
\usepackage{aas_macros}
\usepackage[top=1.0in,bottom=1.0in,left=1.0in,right=1.0in,headsep=10pt,letterpaper]{geometry} 
\usepackage{xcolor} 
\definecolor{ocre}{RGB}{52,177,201} 

\usepackage{avant} 
\usepackage{mathptmx} 

\usepackage{microtype} 
\usepackage[utf8]{inputenc} 
\usepackage[T1]{fontenc} 

\usepackage{titlesec} 

\usepackage{graphicx} 
\graphicspath{{Pictures/}{figures/cosmo/}} 

\usepackage{lipsum} 

\usepackage{tikz} 

\usepackage[english]{babel} 

\usepackage{enumitem} 
\setlist{nolistsep} 

\usepackage{booktabs} 

\usepackage{eso-pic} 

\usepackage{titletoc} 

\contentsmargin{0cm} 
\titlecontents{chapter}[1.25cm] 
{\addvspace{15pt}\large\sffamily\bfseries} 
{\color{ocre!60}\contentslabel[\Large\thecontentslabel]{1.25cm}\color{ocre}} 
{}
{\color{ocre!60}\normalsize\sffamily\bfseries\;\titlerule*[.5pc]{.}\;\thecontentspage} 
\titlecontents{section}[1.25cm] 
{\addvspace{5pt}\sffamily\bfseries} 
{\contentslabel[\thecontentslabel]{1.25cm}} 
{}
{\sffamily\hfill\color{black}\thecontentspage} 
[]
\titlecontents{subsection}[1.25cm] 
{\addvspace{1pt}\sffamily\small} 
{\contentslabel[\thecontentslabel]{1.25cm}} 
{}
{\sffamily\;\titlerule*[.5pc]{.}\;\thecontentspage} 
[]

\titlecontents{lsection}[0em] 
{\footnotesize\sffamily} 
{}
{}
{}

\titlecontents{lsubsection}[.5em] 
{\normalfont\footnotesize\sffamily} 
{}
{}
{}

\usepackage{fancyhdr} 

\fancypagestyle{plain}{\fancyhead{}} 

\makeatletter
\renewcommand{\cleardoublepage}{
\clearpage\ifodd\c@page\else
\hbox{}
\vspace*{\fill}
\thispagestyle{empty}
\newpage
\fi}

\usepackage{amsmath,amsfonts,amssymb,amsthm} 

\newtheoremstyle{ocrenumbox}
{0pt}
{0pt}
{\normalfont}
{}
{\small\bf\sffamily\color{ocre}}
{\;}
{0.25em}
{\small\sffamily\color{ocre}\thmname{#1}\nobreakspace\thmnumber{\@ifnotempty{#1}{}\@upn{#2}}
\thmnote{\nobreakspace\the\thm@notefont\sffamily\bfseries\color{black}---\nobreakspace#3.}} 

\newtheoremstyle{blacknumex}
{5pt}
{5pt}
{\normalfont}
{} 
{\small\bf\sffamily}
{\;}
{0.25em}
{\small\sffamily{\tiny\ensuremath{\blacksquare}}\nobreakspace\thmname{#1}\nobreakspace\thmnumber{\@ifnotempty{#1}{}\@upn{#2}}
\thmnote{\nobreakspace\the\thm@notefont\sffamily\bfseries---\nobreakspace#3.}}

\newtheoremstyle{blacknumbox} 
{0pt}
{0pt}
{\normalfont}
{}
{\small\bf\sffamily}
{\;}
{0.25em}
{\small\sffamily\thmname{#1}\nobreakspace\thmnumber{\@ifnotempty{#1}{}\@upn{#2}}
\thmnote{\nobreakspace\the\thm@notefont\sffamily\bfseries---\nobreakspace#3.}}

\newtheoremstyle{ocrenum}
{5pt}
{5pt}
{\normalfont}
{}
{\small\bf\sffamily\color{ocre}}
{\;}
{0.25em}
{\small\sffamily\color{ocre}\thmname{#1}\nobreakspace\thmnumber{\@ifnotempty{#1}{}\@upn{#2}}
\thmnote{\nobreakspace\the\thm@notefont\sffamily\bfseries\color{black}---\nobreakspace#3.}} 
\makeatother

\newcounter{dummy}
\numberwithin{dummy}{section}
\theoremstyle{ocrenumbox}
\newtheorem{theoremeT}[dummy]{Theorem}

\newtheorem{exerciseT}{Exercise}[chapter]
\theoremstyle{blacknumex}
\newtheorem{exampleT}{Example}[chapter]
\theoremstyle{blacknumbox}

\newtheorem{definitionT}{Definition}[section]
\newtheorem{corollaryT}[dummy]{Corollary}
\theoremstyle{ocrenum}

\RequirePackage[framemethod=default]{mdframed} 

\newmdenv[skipabove=7pt,
skipbelow=7pt,
backgroundcolor=black!5,
linecolor=ocre,
innerleftmargin=5pt,
innerrightmargin=5pt,
innertopmargin=5pt,
leftmargin=0cm,
rightmargin=0cm,
innerbottommargin=5pt]{tBox}

\newmdenv[skipabove=7pt,
skipbelow=7pt,
rightline=false,
leftline=true,
topline=false,
bottomline=false,
backgroundcolor=ocre!10,
linecolor=ocre,
innerleftmargin=5pt,
innerrightmargin=5pt,
innertopmargin=5pt,
innerbottommargin=5pt,
leftmargin=0cm,
rightmargin=0cm,
linewidth=4pt]{eBox}	

\newmdenv[skipabove=7pt,
skipbelow=7pt,
rightline=false,
leftline=true,
topline=false,
bottomline=false,
linecolor=ocre,
innerleftmargin=5pt,
innerrightmargin=5pt,
innertopmargin=0pt,
leftmargin=0cm,
rightmargin=0cm,
linewidth=4pt,
innerbottommargin=0pt]{dBox}	

\newmdenv[skipabove=7pt,
skipbelow=7pt,
rightline=false,
leftline=true,
topline=false,
bottomline=false,
linecolor=gray,
backgroundcolor=black!5,
innerleftmargin=5pt,
innerrightmargin=5pt,
innertopmargin=5pt,
leftmargin=0cm,
rightmargin=0cm,
linewidth=4pt,
innerbottommargin=5pt]{cBox}

\makeatletter
\renewcommand{\@seccntformat}[1]{\llap{\textcolor{ocre}{\csname the#1\endcsname}\hspace{1em}}}
\renewcommand{\section}{\@startsection{section}{1}{\z@}
{-2ex \@plus -1ex \@minus -.2ex}
{1ex \@plus.1ex }
{\normalfont\large\sffamily\bfseries}}
\renewcommand{\subsection}{\@startsection {subsection}{2}{\z@}
{-2ex \@plus -0.1ex \@minus -.2ex}
{0.5ex \@plus.2ex }
{\normalfont\sffamily\bfseries}}
\renewcommand{\subsubsection}{\@startsection {subsubsection}{3}{\z@}
{-2ex \@plus -0.1ex \@minus -.2ex}
{.2ex \@plus.2ex }
{\normalfont\small\sffamily\bfseries}}
\renewcommand\paragraph{\@startsection{paragraph}{4}{\z@}
{-2ex \@plus-.2ex \@minus .2ex}
{.1ex}
{\normalfont\small\sffamily\bfseries}}

\usepackage{hyperref}
\hypersetup{hidelinks,backref=true,pagebackref=true,hyperindex=true,colorlinks=false,breaklinks=true,urlcolor= ocre,bookmarks=true,bookmarksopen=false,pdftitle={Title},pdfauthor={Author}}

\newcommand{\thechapterimage}{}
\newcommand{\chapterimage}[1]{\renewcommand{\thechapterimage}{#1}}

\def\thechapter{\arabic{chapter}}
\def\@makechapterhead#1{
\thispagestyle{empty}
{\centering \normalfont\sffamily
\ifnum \c@secnumdepth >\m@ne
\if@mainmatter
\startcontents
\begin{tikzpicture}[remember picture,overlay]
\node at (current page.north west)
{\begin{tikzpicture}[remember picture,overlay]
\node[anchor=north west,inner sep=0pt] at (0,0) {\includegraphics[width=\paperwidth]{\thechapterimage}};
\draw[anchor=west] (5cm,-9cm) node [rounded corners=20pt,fill=ocre!10!white,text opacity=1,draw=ocre,draw opacity=1,line width=1.5pt,fill opacity=.6,inner sep=12pt]{\huge\sffamily\bfseries\textcolor{black}{\thechapter. #1\strut\makebox[22cm]{}}};
\end{tikzpicture}};
\end{tikzpicture}}
\par\vspace*{230\p@}
\fi
\fi}

\def\@makeschapterhead#1{
\thispagestyle{empty}
{\centering \normalfont\sffamily
\ifnum \c@secnumdepth >\m@ne
\if@mainmatter
\begin{tikzpicture}[remember picture,overlay]
\node at (current page.north west)
{\begin{tikzpicture}[remember picture,overlay]
\node[anchor=north west,inner sep=0pt] at (0,0) {\includegraphics[width=\paperwidth]{\thechapterimage}};
\draw[anchor=west] (5cm,-6cm) node [rounded corners=20pt,fill=ocre!10!white,fill opacity=.6,inner sep=12pt,text opacity=1,draw=ocre,draw opacity=1,line width=1.5pt]{\LARGE\sffamily\bfseries\textcolor{black}{#1\strut\makebox[22cm]{}}};
\end{tikzpicture}};
\end{tikzpicture}}
\par\vspace*{130\p@}
\fi
\fi
}
\makeatother

\usepackage{graphicx}
\graphicspath{{./figures/}}
\usepackage{appendix}
\usepackage{latexsym,amsmath,amsfonts,amssymb,booktabs}
\usepackage[font=small]{caption}
\usepackage{slashed,upgreek,amscd,cancel,tensor,color}
\usepackage{adjustbox}
\usepackage[numbers,sort&compress,square]{natbib}
\usepackage{epsfig,latexsym}
\usepackage{url}
\numberwithin{equation}{section}
\usepackage{doi}
\usepackage{subcaption}
\usepackage{mathtools}
\usepackage{upgreek}
\usepackage{stfloats}
\usepackage{afterpage}
\usepackage{multirow}

\begin{document}


\chapterimage{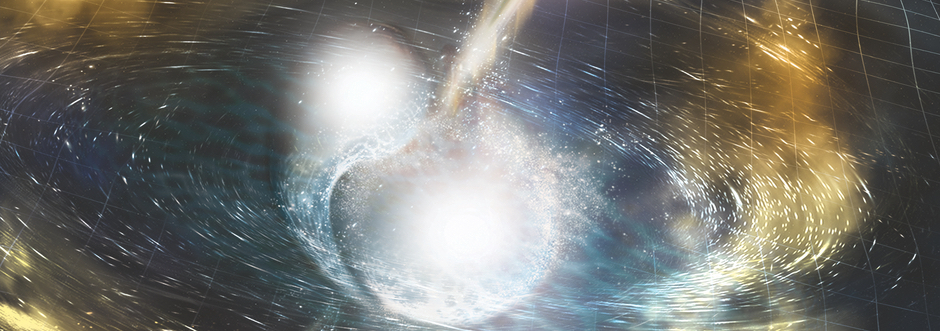} 
\chapter*{\large Multimessenger Universe with Gravitational Waves from Binaries}

\begin{center}
\Large
\textbf{Astro2020 Science White Paper} \linebreak


MULTIMESSENGER UNIVERSE \linebreak with GRAVITATIONAL WAVES from BINARY SYSTEMS \linebreak

\end{center} 

\normalsize
 

\noindent \textbf{Thematic Areas:} 
\begin{itemize}
\item Formation and Evolution of Compact Objects 
\item Stars and Stellar Evolution 
\item Multi-Messenger Astronomy and Astrophysics
\end{itemize}


\vspace{0.75cm}

\noindent \textbf{Principal Author}: 
\newline
\vspace{-0.5cm}

 Name:	B.S. Sathyaprakash
 \newline
\vspace{-0.5cm}

Institution: The Pennsylvania State University
 \newline
\vspace{-0.5cm}

Email: bss25@psu.edu
 \newline
\vspace{-0.5cm}

Phone: +1-814-865-3062 
 \newline

 \noindent 
\textbf{Lead Co-authors:} 
Matthew Bailes (Swinburne U.), 
Mansi M. Kasliwal (Caltech), 
Samaya Nissanke (U. of Amsterdam), 
Shreya Anand (Caltech), 
Igor Andreoni (Caltech), 
Monica Colpi (U.\ of Milano -- Bicocca), 
Michael Coughlin (Caltech), 
Evan Hall (MIT), 
Vicky Kalogera (Northwestern U.), 
Dan Kasen (UC Berkeley),  
Alberto Sesana (U. Birmingham)

\vspace{0.5cm}

 \noindent 
Click here for \href{https://docs.google.com/spreadsheets/d/1uNKEW77Fm-_nc21_3jSOX-P4ecRxsxA5UgDOD4bkG9Y/edit#gid=755774700}{\bf other co-authors and supports}




\clearpage

\section*{Multimessenger Universe with Gravitational Waves from Binaries}
The discovery of GW170817 \cite{TheLIGOScientific:2017qsa} was
a watershed moment in astronomy.
Gravitational wave (GW) and electromagnetic (EM)
observations of this event provided incontrovertible
evidence that binary neutron star (BNS) mergers are connected to short
gamma-ray bursts \cite{Goldstein:2017mmi} and the precise optical localization \cite{Coulter:2017} 
unveiled that these are prolific sites of heavy element
nucleosynthesis. Furthermore, they showed that to an outstanding
accuracy the speed of gravitational waves is identical to
the speed of light and allowed the first measurement of the
Hubble constant using GW standard sirens \citep{Abbott:2017xzu,2018Nature.561..355M}, 
ushering in a new era in cosmology. Observations of the event in the entire EM
window \cite{abbott17_MM} have accumulated a treasure trove
of data that will have a lasting impact on our understanding
of some of the most energetic phenomena in the Universe and
matter in extreme environs.

\begin{tcolorbox}[standard jigsaw,colframe=ocre,colback=ocre!10!white,opacityback=0.6,coltext=black]
Future GW detector networks and EM observatories will provide a unique
opportunity to observe the most luminous events in the
Universe involving matter in extreme environs. The observations
will address some of the key questions in physics and astronomy:
\begin{itemize}[leftmargin=*]
\item {\bf Formation and Evolution of Compact Binaries.} 
How do double neutron star and neutron star-black hole
binaries form and evolve; what are their demographics,
merger rates, and mass and spin distributions as a function
of redshift? 
\item {\bf Sites of Formation of Heavy Elements.} What is
the role of neutron star mergers in the production of heavy
elements in the Universe? Are they able to explain
abundances in the Solar System and stars?
\item {\bf Jet physics.} 
What is the physics of central engines in mergers, and how do they relate to short gamma-ray bursts? 
How do the jet properties vary with progenitor binary parameters?
\item{\bf Multi-band GW Astronomy.} What can joint observations by the Laser
Interferometer Space Antenna and the 3G network 
tell us about the origin and evolution of black holes?
\end{itemize}
\end{tcolorbox}

\noindent{\bf Capabilities of Next Generation Detector Networks:}
The next generation of GW detectors (3G, see Table
\ref{table:localize_BNS_100Myr}) will compile surveys of the
Universe for close binary coalescence events in which one of
the companions is a neutron star and the other is either a
stellar mass black hole or also a neutron star.  
The Table shows the capability
of a third generation detector network (3G) compared to the
current network of advanced detectors at their design
sensitivity.  
\vskip-10pt
\begin{table}[h!]
\parbox{0.425\linewidth}{
For this simulation, source redshifts were sampled from a merger redshift distribution of binary
neutron stars, assuming the Madau-Dickinson star formation
rate, with an exponential time delay between formation and
merger with e-fold time of 100 Myr (see
\cite{Vitale:2018yhm}) and a local co-moving BNS merger rate
of 1000 Gpc$^{-3}$ yr$^{-1}$.  It is clear that the 3G
network will provide ample opportunity for EM follow-up of
BNS mergers.  Key science questions addressed by the detected 
population in the 3G era is very rich and diverse.  Localizing the EM
}
\hfill
\parbox{.53\linewidth}{
\small
\begin{tcolorbox}[standard jigsaw,colframe=ocre,colback=ocre!10!white,opacityback=0.6,coltext=black]
\vskip-5pt
\caption
{
\small Expected detections per year ($N$), number
detected with a resolution of $< 1$, $<10$ and
$<100$ sq.\,deg. ($N_1,$ $N_{10}$ and $N_{100}$, 
respectively) and median localization error ($M$ in sq.\,deg.), in a network
consisting of LIGO-Hanford, LIGO-Livingston and Virgo (HLV),
HLV plus KAGRA and LIGO-India (HLVKI) and 1 Einstein
Telescope and 2 Cosmic Explorer detectors (1ET+2CE). 
}
\label{table:localize_BNS_100Myr}
\centering
\vskip-5pt
\begin{tabular}{lccccc}
\hline
 Network      &   $N$    &   $N_1$ & $N_{10}$ & $N_{100}$ & $M$ \\
\hline
 HLV          &       48 &       0 &      16  &      48   & 19 \\
 HLVKI        &       48 &       0 &      48  &      48   &  7 \\
 1ET+2CE      &    990k  &     14k &    410k  &    970k   & 12 \\
\hline
\end{tabular}
\end{tcolorbox}
}
\end{table}
\vskip-14pt
\noindent counterpart to
such events will allow us to characterize matter in extreme
environments. The redshift of the host galaxy enables
cosmological applications, whilst the sub-arcsecond
localization of the kilonova provides information about the
nucleosynthesis, environment of the event, jet physics and
formation scenarios.  

\subsection*{Demographics of Compact Binary Mergers}
A key question about compact binary mergers is their
demographics, as this could reveal their formation mechanism.
Localization of merger events to less than galactic scales
($\sim 30$ kpc) is essential to unambiguously infer
associations of mergers with their host galaxies.  Without
an EM counterpart the vast majority of events will have
error boxes that greatly exceed the typical radii of
potential host galaxies. The merger
fraction split between early type and star-formation
galaxies will provide a fascinating insight into the
fraction of mergers that are created with short
gravitational ``fuses'' \citep{Chaurasia:2005aq} that are comparable to the
evolutionary timescales of massive stars and those that
extend out to a Hubble time. Their locations \citep{Bloom:1998zi,Abbott:2017ntl} within the
hosts will give insights into the kick velocities imparted
to the binaries during their supernova explosions.

EM follow-up of BNS mergers will be critical in pinning down
host galaxies.  Binary black hole (BBH) mergers, not believed to
produce any EM counterparts, will not be resolved well
enough to unambiguously identify their hosts.  The situation is more
optimistic for neutron star-black hole (NSBH) mergers.
Theoretical predictions suggest that when the mass ratio is
not too extreme depending on the BH spin, conditions could be favorable for the
creation of an accretion disk around that
might rival the absolute visual magnitude of the GW170817
kilonova, and, therefore, be detectable out to $z=0.5$ in the
reddest filters.  
If such mergers occur in the globular cluster cores 
it will be difficult to identify host clusters much beyond 
Virgo, and those in Virgo do not require a 3G GW detector 
for discovery.

In the modern paradigm, galaxies are assembled by the merger of
smaller proto-galaxies and star formation peaks near $z\sim 2$
\citep{madau2014}. Identification of kilonovae beyond $z\sim 0.5$
requires hour-long integrations on 8m class facilities like LSST or
Subaru and therefore determining the host galaxies of BNS mergers near
the peak of star formation will not be routine in the absence of a
gamma-ray burst jet pointing towards the Earth, even with ELTs.
Nevertheless, at redshifts $z<0.5$ 3G detectors will work in concert
with astronomy facilities to enable thousands of host galaxy
identifications from BNS and NSBH mergers thanks to the identification
of a kilonova. At larger distances, the identification will be
possible only through the detection of an associated gamma-ray burst
afterglow,  which can be much more luminous than a kilonova if the jet
is directed towards the Earth.

\subsection*{Nucleosynthesis in Binary Neutron Star Mergers}
A long standing puzzle in astrophysics is how the elements
heavier than iron came into being. About half of these
elements are believed to have been created by a process of rapid
neutron capture (the ``r-process"), but it is unclear which
astrophysical sites are the main contributors. Neutron star
mergers have long been proposed as a possible site\cite{Lattimer:1974slx}. 
GW170817 and its associated thermal EM counterpart provided the first
direct identification of a prolific site of r-process
nucleosynthesis \cite{abbott17_MM}.
However, determining the degree to which
BNS mergers contribute to cosmic chemical abundance and
evolution will require a more extensive determination of the
rates, locations, timescales, and nucleosynthetic yields of
the various types of merger events. Even the basic question of
whether all three r-process abundance peaks were synthesized by GW170817 is
debated \cite{Rosswog:2017sdn,Kasliwal:2018fwk}.

Heavy elements can be synthesized in BNS or NSBH mergers
when clouds of neutron-rich material are expelled, either
dynamically during the merger or later in the form of winds blown off the
remnant accretion disk. The subsequent
radioactive decay of the freshly synthesized elements powers
a thermal optical/infrared transient known as a “kilonova”.
Theoretical modeling has demonstrated how measurements of
the brightness and color of the kilonovae are diagnostic of
both the total mass of r-process elements and the relative
abundance of lighter to heavier elements \cite{Kasen:2017}. 

Whereas historical studies of chemical evolution have relied
on observing fossil traces of r-process elements mixed into
old stars, multi-messenger observations (MMOs) provide the
unique opportunity to study heavy element formation
at its production site and to determine how the initial
conditions of an astrophysical system map to the final
nucleosynthetic outcome.  Answering the basic question of
the extent to which BNS and NSBH mergers are the dominant site of
r-process production will require MMOs of a large sample of
events. GW measurements would pin down the rate of mergers
and the binary properties, such as the binary type 
(BNS or NSBH), companion masses, the merged
remnant lifetime and the spin-orbit alignment,
while optical/infrared photometry of the associated
kilonovae would determine the average r-process yields.
Detailed infrared characterization would probe the relative
abundance distribution and how similar or different it is
from the solar abundance distribution of heavy elements.
These observations would also
illuminate the key physics driving the r-process and
kilonova, such as the equation of state of dense
matter, the fundamental interactions of neutrinos and the
magneto-hydrodynamics of accretion.

Statistical studies of MMOs will reveal how r-process
production in BNS and NSBH mergers depends on host galaxy
type, location and redshift, allowing us to piece together
the history of when and where the heavy elements were formed
over cosmic time.  Such studies can determine the
distribution of delay times between star formation and
merger, thereby addressing whether
some BNS and NSBH mergers occurred  promptly enough to
explain the enrichment of the oldest metal poor stars and
the extent to which compact 
binaries receive strong kicks that move them within, or expel
them from, their host galaxies, a factor that is important
for understanding whether mergers can explain
the unusually high r-process enhancement seen in some dwarf
galaxies \citep{Ji:2015wzg}.

In addition to discovering BNS and NSBH mergers beyond the 
peak of star formation, 3G detectors, because of their wide band sensitivity,
will track the full inspiral, merger and ringdown signal. This 
could enable exquisite measurements of the intrinsic masses prior to 
coalescence, and determine companion spins and the nature of the 
remnant, key parameters to fully determine the dynamics of the
merger, the nature of the relic star, and the type of debris responsible for
panchromatic emission of radiation. State of the art numerical
magneto-hydro-dynamical simulations will provide key insight into the geometry 
and physical state of the debris, in the form of ejecta, winds and discs that can be 
used to model the EM signal. Thus, the combination of information derived independently from the
EM and GW signals will be immensely powerful to build a complete, self-consistent astrophysical picture.

In the era of 3G detectors, optical kilonovae will be
detectable by LSST out to 3 Gpc and infrared
characterization photometrically by WFIRST/Euclid and
spectroscopically by JWST/GMT/TMT/E-ELT would be out to 1
Gpc (z$<$0.2). In summary, building the necessary sample size of a
thousand kilonovae with detailed ultraviolet-optical-infrared follow-up to probe
the nucleosynthesis and ejecta properties is a realistic goal in the 3G era.

\subsection*{Jet Physics in BNS and NSBH Mergers}
Relativistic explosions and compact-object mergers can
generate collimated, energetic jets of material and
radiation. Our understanding of jet physics thus far comes
from studies of gamma-ray bursts, active galactic nuclei and X-ray binaries.
Multi-messenger observations provide an entirely new
perspective on this topic.  

The panchromatic study of GW170817 revealed that there was
both a wide-angle mildly relativistic cocoon 
\citep{Nakar:2018, Kasliwal:2017ngb, Mooley:2017enz} 
as well as a narrow ultra-relativistic jet 
\citep{2018Nature.561..355M, 2018arXiv180800469G, Margutti:2018, Lamb:2018qfn}. 
This was not seen in previous
studies of cosmological short-hard gamma-ray bursts.
Combining the EM and GW allowed to directly constrain system
parameters with unprecedented precision. 
GW170817 opened up many questions for future
events to answer. Specifically, what is the connection to
the class of cosmological short hard gamma-ray bursts? Does
a wide-angle mildly relativistic cocoon always accompany a
BNS merger? Does the jet always successfully escape the
cocoon or is it sometimes choked? How do the observed
jet properties vary as a function of viewing angle, mass
ratio, hypermassive neutron star lifetime, remnant spin, and ejecta mass?
 Do mergers produce prompt EM signals?
What is the distribution of the time delays between the EM
and GW signal arrival times? What are the
characteristics of a jet from a NSBH merger? With the first 
census of BNS and NSBH coalescences, and full GW and EM
coverage of the signals, joint multi-messenger Bayesian parameter
inference will be key in understudying the physical origin of jets,
ubiquitous around relativistic sources \cite{Bauswein:2017,Hinderer:2018,Coughlin:2018,Radice:2018}.
For the first time, a direct measurement of the BH spin in a source
emitting a collimated jet, will enable to establish the close correlations
between the jet power, the spin and the inflow rate from the debris disk,
which determines the conditions for launching the jet.

The 3G GW network combined with new, powerful EM
facilities can further revolutionize our
understanding of the physics of jets.  3G network will enable the
detection of neutron-star mergers out to redshifts of $z\sim
10.$ Even with the planned upgrades, we are limited by
the sensitivity of gamma-ray, X-ray and
radio telescopes to study jet physics to only out to 500\,Mpc.
To build a sample large enough to map the full parameter
space, we would need $\sim $ thousand events localized to
better than few sq deg. This is a realistic goal with the
proposed 3G network. 

\subsection*{Cosmology}
Joint GW and EM observations provide a completely independent tool
for measuring the dynamics of the universe and to constrain cosmological parameters 
such as the Hubble parameter, dark matter and dark energy densities, 
and the equation of state of dark energy 
\citep{Schutz:1986gp,Abbott:2017xzu,Sathyaprakash:2009xt,Zhao:2010sz,Cai:2016sby}. 
It is estimated that about 10 compact BNS or NSBH 
mergers with EM counterparts are required to
reach $H_0$ measurements at the 5\% level, and 200 to reach
1\% \citep{2010ApJ...725..496N,Chen:2017rfc,2019PhRvL.122f1105F}. 
While BNS events are promising based on GW170817,
NSBH mergers, due to precession of the orbital plane because of spin-orbit coupling,
can break the degeneracy between the orbital inclination
and luminosity distance, and provide accurate distance
measurements \citep{Vitale:2018wlg}. In addition, EM observations could also break this degeneracy \citep{Hotokezaka:2018}. 
There is significant potential in statistical methods as
well, where sources without EM counterparts are
combined with galaxy catalogs to make inferences \citep{DelPozzo:2011yh}. For
example, 3G detectors will localize some
BBHs within a volume where on average only one
galaxy is present \citep{Vitale:2018nif,2017ApJ...848L..16S}, although the method is 
limited by the
peculiar velocity at the redshift of interest and the distance
uncertainty from GW observations ($\approx 1$\%).

\subsection*{Multi-band GW astronomy with LISA}

Joint observations of GW events by the Laser
Interferometer Space Antenna (LISA) at milli-Hertz
frequencies and 3G detectors at audio frequencies could
maximise their science potential. If LISA had been observing
in 2010, it would have detected GW150914 years
before it was observed by LIGO
\citep{2016PhRvL.116w1102S}. Indeed, LISA will see up to
thousands of stellar-mass BBH mergers  of $M>20-30 M_\odot,$
up to $z\approx0.3$
\citep{2016PhRvL.116w1102S,2016MNRAS.462.2177K}. A small
fraction of them will sweep across the detector band within
few years that will eventually be detected by ground-based
detectors. The benefit of multi-band observations of such
events will be quite significant.

\begin{table}
\parbox{1.0\linewidth}{
\begin{tcolorbox}[standard jigsaw,colframe=ocre,colback=ocre!10!white,opacityback=0.6,coltext=black]
\small
\caption{Present ($P$) and future ($F$) electromagnetic facilities that are able to observe faint/distant counterparts to GWs. Detection Limit ({\bf DL}, 1 hr exposure time) for UV, optical, and near-IR facilities are expressed in AB magnitudes, for X-rays in $10^{-16}\,\rm erg\,s^{-1}\,cm^{2},$ and for radio in $\mu$Jy.  Distance reach ({\bf D} in Mpc) of facilities for GW170817-like events are shown.}
\vskip-5pt
\label{tab:facilities}
\parbox{.50\linewidth}{
	\begin{tabular}{cllr} 
		\hline
		\hline
	    & {\bf Facility} & {\bf DL} & {\bf D} \\
        \hline \\
         Gamma-rays 
         &{\it Fermi} $P$ & S/N 5 & 80 \\ 
         & AMEGO $F$ & S/N 5 & 130 \\ 
        \hline
         &  {\it Swift} $P$ & S/N 5 & $\sim$80 \\ 
         & {\it Chandra} $P$ & 30 & 150 \\
         X-rays
         & ATHENA $F$ & 3 & 480 \\
         & {\it Lynx} $F$ & 6  & 450 \\%
         & STROBE-X $F$ & S/N 5 & 120 \\
        \hline UV 
        &  HST (im) $P$ & 26 & 2000 \\
        &  HST (spec) $P$ & 23 & 400 \\
        \hline
        Optical 
        & Subaru $P$ & 27 & 3200 \\
        Imaging 
        & LSST $F$ & 27 & 3200 \\ \\
        \hline
        \end{tabular}
        }
        \hskip30pt
\parbox{.50\linewidth}{
	\begin{tabular}{cllr} 
        \hline
        &   Keck/VLT $P$ & 23 & 500 \\
        & Gemini Obs. $P$ & 23 & 500 \\
        Optical
        & GMT $F$ & 25 & 1265 \\
        Spec.
        & TMT $F$ & 25.5 & 1592 \\
        & E-ELT $F$ & 26 & 2005 \\
        \hline
        Infrared 
        &  WFIRST $F$ & 27.5 & 4800  \\
        Imaging
        & Euclid $F$ & 25.2 & 1700 \\
        \hline
        & Keck/VLT & 21.5 & 481  \\
        Infrared 
        & GMT $F$ & 23.5 & 762 \\
        Spec.
        & TMT $F$ & 24 & 960 \\
        & E-ELT $F$ & 24.5 & 1208 \\
        \hline
        \multirow{4}{*}{Radio} 
        & VLA (S) $P$ & 5 & 91 \\
        &ATCA (CX) $P$ & 42 & 51 \\
        &ngVLA (S) $F$ & 1.5 & 353 \\
        &SKA-mid (L) $F$ & 0.72 & 634 \\
		\hline
	\end{tabular}
        }
        \end{tcolorbox}
        }
        \vskip-5pt
\end{table}     

LISA would provide a precise measurement of the system's
eccentricity (to a precision of $\Delta{e}<0.001$
\citep{2016PhRvD..94f4020N}), sky localization to 0.1
sq.~deg and time to coalescence within few seconds, several
weeks prior to coalescence \citep{2016PhRvL.116w1102S}. This
will help point an EM telescope in the right direction {\em
before} the merger, providing a much deeper coverage from
radio to gamma-ray than what might be possible without any
early warning alert. This could also allow real
time optimization of 3G detectors to tune their sensitivity
to observe the ringdown signal, thus enhancing the potential
of BH spectroscopy \citep{Tso:2018pdv}. On
the other hand, one can use the information extracted by 3G
network to dig out sub-threshold LISA events
\citep{Wong:2018uwb}. From an astrophysical
standpoint, the eccentricity information delivered by LISA
can be combined with the spin measurement obtained from 3G
detectors to better constrain different formation channels
\citep{2016ApJ...830L..18B,2017MNRAS.465.4375N,Samsing:2018ykz}.
Multi-band observations will also facilitate tests of GR
\citep{2016PhRvL.117e1102V} by enhancing the sensitivity to
specific deviations arising in the long adiabatic inspiral
as predicted, for example, from dipole radiation
\citep{2016PhRvL.116x1104B}.  

The future of multi-band GW astronomy lives in the observation of 
intermediate-mass black holes \cite{2010ApJ...722.1197A,Amaro-Seoane:2018gbb},
the most elusive sources that might form at the center of dense 
clusters as the end-product of runaway collisions
\citep{2004Natur.428..724P}, or result from the death of
very massive, low metallicity stars
\citep{2003ApJ...591..288H}. 
Multi-band GW astronomy will also be realized by complementary
observations of different samples of the same population of
sources. Seed black holes are a case in point
\citep{2010A&ARv..18..279V}:  While LISA will be sensitive
to mergers of $M \ge 10^3 M_\odot$ binaries out a redshift
of $z>20$, 3G detectors can access $M \sim 100 M_\odot$
population at comparable redshifts. If 3G detectors see a
$\sim 100 M_\odot$ merger at high $z$, one question that
arises is whether those are the seeds of massive black holes
(MBHs) or a different population that will not evolve into
MBHs. Multi-band GW observations will likely
solve this issue by quantifying the continuity between the
two populations. 


\chapterimage{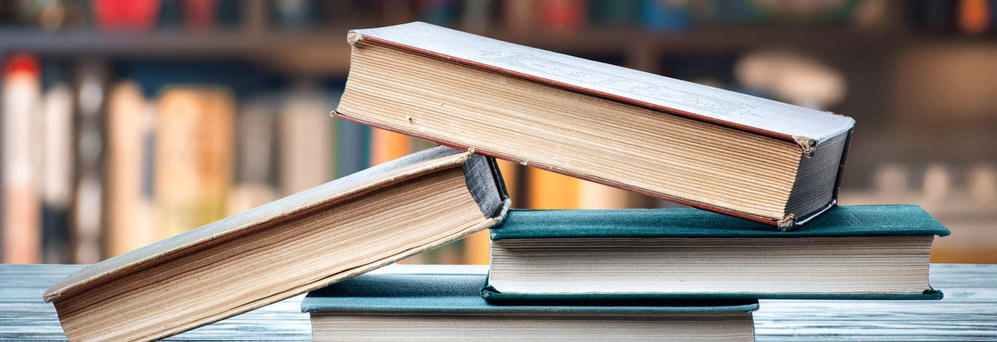} 
\bibliographystyle{utphys}
\bibliography{wp-all,3g}

\end{document}